\documentclass[aps,prd,twocolumn,superscriptaddress,showpacs,
amsmath,amssymb,nofootinbib,hyperref,
eqsecnum,
longbibliography,
 preprintnumbers]{revtex4-2}
\usepackage{graphicx}
\usepackage{ulem}
\usepackage{epstopdf}
\usepackage{color}
\usepackage{enumerate}

\newcommand{\be}{\begin{equation}}
\newcommand{\ee}{\end{equation}}
\newcommand{\ba}{\begin{eqnarray}}
\newcommand{\ea}{\end{eqnarray}}

\newcommand{\beq}{\begin{equation}}
\newcommand{\eeq}{\end{equation}}
\newcommand{\beqa}{\begin{eqnarray}}
\newcommand{\eeqa}{\end{eqnarray}}

\begin{document}

\title{Can Bekenstein’s area law prevail in modified theories of gravity?}

\author{David Kubiz\v{n}{\'a}k}
\email{david.kubiznak@matfyz.cuni.cz}
\affiliation{Institute of Theoretical Physics, Faculty of Mathematics and Physics,
Charles University, Prague, V Hole\v{s}ovi\v{c}k\'ach 2, 180 00 Prague 8, Czech Republic}

\author{Marek Li{\v s}ka}
\email{liska.mk@seznam.cz }
\affiliation{Institute of Theoretical Physics, Faculty of Mathematics and Physics,
Charles University, Prague, V Hole\v{s}ovi\v{c}k\'ach 2, 180 00 Prague 8, Czech Republic}

\date{November 14, 2023}

\begin{abstract}
According to Bekenstein's area law, the black hole entropy is identified holographically -- with one quarter of the horizon area. However, it is commonly believed that such a law is only valid in Einstein's theory and that higher curvature corrections generically give rise to its modifications. This is for example the case of black holes in Lovelock gravities, or their four-dimensional cousins in the recently discovered 
4D scalar-tensor Gauss--Bonnet gravity where one naively `finds' (classical) logarithmic corrections to the Bekenstein's law. In this Letter we argue that such logarithmic corrections originate from ignoring the shift symmetry of the 4D Gauss--Bonnet gravity. When this symmetry is properly taken into account, there is no longer any departure from the area law in this theory. Moreover, the first law remains valid upon modifying the black hole temperature, which 
can be derived via the Euclidean grand canonical ensemble (Brown--York) procedure, but is no longer given by the surface gravity. Interestingly, we show that upon similar modification of the black hole temperature the area law can also prevail for black holes in higher-dimensional Lovelock gravities.  
 \end{abstract}


\maketitle

\section{Introduction}

It is now more than 50 years since Bekenstein analyzed what happens upon throwing a cup of tea into a black hole. Based on this, he proposed that black holes should carry entropy proportional to their horizon area \cite{Bekenstein:1973}. 
After the constant of proportionality was fixed by Hawking's calculation of the black hole temperature \cite{Hawking:1975}, the Bekenstein area law states that the black hole entropy is given by {\it  one quarter of the horizon area} (using in what follows the natural units where  $G_N=c=\hbar=k_B=1$):
\be\label{Alaw}
S=\frac{\mbox{Area}}{4}\,.
\ee 
With these identifications of entropy and temperature, the laws of black hole mechanics \cite{Bardeen:1973} became the laws of black hole thermodynamics.

However, while the thermodynamic description of black holes is universal -- remaining valid beyond Einstein's gravity -- it is commonly believed that the black hole entropy depends on gravitational dynamics and will in general differ from the area law \cite{Wald:1994}.\footnote{In fact, even within Einstein's gravity it was believed for a long time that the area law gets modified for the so called Taub--NUT solutions, with the modification accounting for the existence of the so called Misner strings. However, as shown recently, a more natural interpretation of the first law is the one where the area law remains preserved and the Misner strings have their own associated thermodynamic charges, e.g. \cite{BallonBordo:2019vrn}.} A canonical example is the Gauss--Bonnet gravity, and more generally the {\it  Lovelock theory} \cite{Lovelock:1971yv} (the most general higher-curvature gravity characterised by second-order equations of motion), where the area law picks up corrections due to `subleading topological densities' (see Eq.~\eqref{SJacobson} below). In particular, taking a specific (singular) limit of Einstein--Gauss--Bonnet gravity to four dimensions, upon which one recovers a certain scalar--tensor theory known as {\it  4D Einstein--Gauss--Bonnet gravity} \cite{Glavan:2019inb, Lu:2020iav,Fernandes:2020nbq, Hennigar:2020lsl},  
the modifications naively yield classical {\it  logarithmic corrections} to the area law. 
  
In this Letter, we argue that the latter conclusion is incorrect, as it ignores the basic scalar field {\it  shift symmetry} of the 4D Gauss--Bonnet gravity. When this symmetry is recovered at the level of the action, by adding an appropriate boundary term, the standard {\it  Noether charge} (Wald entropy) argument \cite{Wald:1994} then shows that the logarithmic corrections to the black hole entropy disappear and the entropy is simply given by the area law \eqref{Alaw}. To satisfy the first law of black hole thermodynamics, the Hawking temperature must then be modified and is no longer proportional to the surface gravity. This temperature modification has been confirmed by a Brown--York construction~\cite{Braden:1990hw, Brown:1992bq} of a Euclidean grandcanonical ensemble~\cite{Liska:2023fdz}.

This is, in a way, similar to the recent proposal that for certain Horndeski theories \cite{Horndeski:1974wa} the temperature gets modified while Bekenstein's law remains valid \cite{Hajian:2020dcq}. In that case, however,  the modification of the black hole temperature was accounted for due to scalar field effects on the propagation of the graviton close to the horizon, which was shown to propagate in the effective geometry, different from the background black hole. 
In our case, the modification of the entropy and the temperature simply arise from the novel boundary term designed to restore the shift symmetry. 
While a change of Wald entropy due to addition of a surface term to the action appears unusual, such a situation is not without a precedent -- it has been known for a long time that by adding the Gauss--Bonnet density in 4D (where it is a surface term) the black hole entropy changes by a universal (horizon topology-dependent) constant \cite{Jacobson:1993xs}. 
Here the effect just becomes more intricate due to the dependence on the scalar field.

In what follows we i) show that the area law prevails for black holes in 4D Gauss--Bonnet gravity 
and their temperature gets accordingly modified when the shift symmetry of the theory is
preserved at the level of the action, and ii) observe that the area law leads to a simplified thermodynamics of `spherical' black holes in higher-dimensional Lovelock gravity 
provided their temperature is similarly  modified.

\section{4d Einstein--Gauss--Bonnet gravity with shift symmetry}
\label{4D EGB}

The idea that Bekenstein's area law may be respected in Lovelock gravity in dimension $d>4$, outlined in the next section, is based on the behavior of entropy in a 4D formulation of Einstein--Gauss--Bonnet gravity. In this theory, we can make a solid case for the survival of the Bekenstein's area law. Then, given the status of this theory as a (singular) limit of higher-dimensional Gauss--Bonnet gravity (a particular case of Lovelock gravity), we speculate in the next section that the area law may be a general feature of Lovelock gravity. 

In 4D, the only possible Lovelock theory is {Einstein's general} relativity, since the higher-curvature topological densities in the {Lovelock action (see \eqref{eq:Loveaction} below)}
either vanish, or, in the case of the 
Gauss--Bonnet term 
\be 
\mathcal{G}=R^2-4R_{ab}R^{ab}+R_{abcd}R^{abcd}\,,
\ee
 become a total divergence, i.e., $\mathcal{G}=\nabla_{a}\mathcal{G}^{a}$. Therefore, they do not affect the gravitational equations of motion. Nevertheless, one can obtain a 4D formulation of Einstein--Gauss--Bonnet gravity by taking a singular limit. While the original idea \cite{Glavan:2019inb} does not quite work \cite{Arrechea:2021}, a consistent theory can be obtained by a conformal trick/certain Kaluza--Klein reduction
\cite{Lu:2020iav,Fernandes:2020nbq, Hennigar:2020lsl}. Both give the same resulting action (in which we also include minimally coupled electromagnetic field):
\ba
I_{\mbox{\tiny EGB}}&=&\frac{1}{16 \pi}\!\int\biggl[R-2\Lambda+\alpha\Bigl(\phi\mathcal{G}+4G^{ab}\nabla_{a}\phi\nabla_{b}\phi \nonumber\\
&&-4(\nabla \phi)^2\Box \phi+2(\nabla \phi)^4\Bigr){-4\pi F_{ab}F^{ab}}\biggr] \sqrt{-\mathfrak{g}}\text{d}^{4}\!x\,, \nonumber \\
\label{action}
\ea
where $\alpha$ denotes the Gauss--Bonnet coupling. The equations of motion obtained from this action contain at most second derivatives of the metric and $\phi$, making this a special case of Horndeski gravity.

Crucially, the equations of motion possess a shift symmetry, i.e., they are not affected by shifting the scalar field by a constant. However, such a shift of the scalar field changes the action by a surface term. This has no effect on the equations of motion, but it affects the construction of the covariant phase space by the Iyer--Wald method \cite{Wald:1994} and, consequently, Wald entropy. In particular, Wald entropy obtained from action \eqref{action} equals
\ba
\nonumber S_{0}&=&\frac{\text{Area}}{4}+\frac{\alpha}{4}\int_{\mathcal{H}}\phi\big(\epsilon^{ab}R-4\epsilon^{c a} R_{c}^{\;\:b} \\
&&\qquad+\epsilon_{cd}R^{ba cd}\big)\epsilon_{ba}\text{d}^2\mathcal{A}\,.
\ea
The second term modifies Bekenstein's area law and seemingly gives rise to a classical logarithmic correction to entropy~\cite{Wei:2020}. However, this term clearly breaks the shift symmetry, as adding a constant to the scalar field changes the entropy by a constant. Given that the equations of motion (and, hence, all the classical physics) are invariant under these shifts, a measurable quantity such as the entropy should not be affected by it. Furthermore, a shift of the scalar field can be used to make the entropy negative.

A natural solution is to add a surface term to action \eqref{action} in order to make it exactly shift symmetric. Then, this symmetry will also be respected by the covariant phase space formalism and Wald entropy. We thus consider the following action:
\begin{equation}
\label{right action}
I_{\mbox{\tiny inv}}=I_{\mbox{\tiny EGB}}-\frac{1}{16\pi 
}\int\nabla_{a}\left(\alpha\phi\mathcal{G}^{a}\right)\sqrt{-\mathfrak{g}}{d}^4\!x\,.
\end{equation}
The equations of motion are unchanged, but the Wald entropy $S$ is now simply given by the area law \eqref{Alaw} (this can be easily verified using a closed expression for $\mathcal{G}^{a}$ valid for spacetimes with Killing symmetries \cite{Padmanabhan:2011}).\footnote{Of the quantities describing a spherical symmetric black hole solution, only the value of the scalar field has a logarithmic behavior. Hence, there can be no shift symmetric expression for black hole entropy containing the logarithmic correction term, see also \cite{Minamitsuji:2023nvh}.

It is easy to verify that adding the surface term \eqref{right action} does not modify the generalized Komar integrals, such as those defining the asymptotic mass.   
However, it may give rise to a novel variational principle (with alternative boundary conditions) whose implications for black hole thermodynamics remain to be clarified, e.g. \cite{Compere:2008us, Odak:2021axr}. We have also checked that, similar to the Einstein gravity, the 4D Gauss--Bonnet York--Gibbons--Hawking term, obtained in \cite{Ma:2020ufk}, does not affect the Wald entropy, and the area law  remains valid even after including such a term.} 

Upon modifying the entropy, we also need to check the validity of the first law of thermodynamics. We do so for the case of electrovacuum, static, spherically symmetric solutions of the theory, which are known analytically. They take the standard spherically symmetric form \cite{Lu:2020iav}
\begin{equation}
{d}s^2=-f{d}t^2+\frac{dr^2}{f}+r^2 d\Omega_2^2\,\quad F=\frac{Q}{r^2}dt\wedge dr\,,
\end{equation}
where $d\Omega_2^2={d}\theta^2+\sin^2\!\theta{d}\phi^2$,
\begin{equation}
\label{f solution}
f=1+\frac{r^2}{2\alpha}\left(1-\sqrt{1+\frac{4}{3}\alpha\Lambda+\frac{8\alpha M}{r^3}-\frac{4\alpha Q^2}{r^4}}\right)\,,
\end{equation}
and the scalar field satisfies
\begin{equation}
\label{phi}
\phi\left(r\right)=\ln\left(\frac{r}{L}\right)\pm\int_{r_+}^{r}\frac{1}{\rho\sqrt{f\left(\rho\right)}}\text{d}\rho\,.
\end{equation}
Here, $M$ and $Q$ stand for the mass and charge, respectively, $\Lambda$ is the bare cosmological constant, $r_+$ is the horizon radius, and $L$ is an arbitrary integration constant with dimensions of length. In this case, the Smarr formula and first law are, respectively
\ba
M&=&2T_{\text{mod}}S{+\Phi Q{-2VP}}
+2\Psi_\alpha\alpha\,,
\label{smarr} \\
\delta M&=&T_{\text{mod}}\delta S+\Phi\delta Q{+V\delta P}
+\Psi_\alpha \delta \alpha\,.
\ea 
We have identified the bare negative cosmological constant with thermodynamic pressure $P=-\Lambda/(8\pi)$, and denoted the conjugate thermodynamic quantity as thermodynamic volume, $V=4\pi r_+^3/3$, see e.g. \cite{Kubiznak:2016qmn}; the last two terms in the first law are only present when $\alpha$ and $\Lambda$ are considered as thermodynamic variables, while their presence is inevitable for the Smarr relation to hold; $\Psi_\alpha=1/(2r_+)$. $\Phi=Q/r_+$ denotes the electric potential on the horizon, and we introduced modified Hawking temperature
\begin{equation}
\label{T mod}
T_{\text{mod}}=T_0\left(1+\frac{2\alpha}{r_+^2}\right)\,,\quad T_0=\frac{\kappa}{2\pi}=\frac{|f'(r_+)|}{4\pi}\,.
\end{equation}
The need to modify Hawking temperature in order to satisfy the first law in Horndeski gravity has been previously proposed \cite{Hajian:2020dcq}. It has been suggested that it occurs due to modified speed of propagation of gravitons due to non-minimal coupling of the scalar field to the metric. Since 4D scalar--tensor Einstein--Gauss--Bonnet gravity is a special case of Horndeski gravity, the temperature modifications can be expected even in our case. All Horndeski theories share an interesting feature. In the Smarr formula \eqref{smarr}, we have two contributions to the 
$T_{\text{mod}}S$ term. One comes from the integral of the Noether charge over the horizon and corresponds to the standard Hawking temperature, the second one is given by the volume integral of the Noether current and gives the correction to it. At the moment, we cannot say whether this is just a mathematical coincidence or it suggests that the temperature corrections are somehow related to what happens far outside the horizon (i.e., {due to the} redshift during the propagation of the radiation to infinity). Notably, a Euclidean grand canonical ensemble approach (Brown--York procedure) allows one to derive the temperature directly by finding stationary points of a `reduced action' \cite{Braden:1990hw}. In this case, we recover precisely the modified temperature we obtained by demanding that the first law of thermodynamics holds. Since this calculation fixes temperature independently of the entropy formula or the first law, it strongly suggests that the temperature is indeed modified. However, it does not seem to provide any clear clue for the physical origin of this modification. We will report on results of this calculation and thermodynamics of 4D scalar--tensor Einstein--Gauss--Bonnet gravity in more detail elsewhere \cite{Liska:2023fdz}.

\section{Area law in Lovelock theories?}  

Let us now turn to (charged) black holes in Lovelock gravity \cite{Lovelock:1971yv}. This is a class of geometric higher curvature {theories of gravity, a natural generalization of Einstein's theory to higher dimensions, 
that give rise to second-order field equations for all metric components.} In $d$ spacetime dimensions, the action, including the electromagnetic part, reads
\begin{equation}
I =  \frac{1}{16\pi} \int {d}^d\!x \sqrt{-\mathfrak{g}}\, \Bigl({\sum_{k=0}^{K} }\hat{\alpha}_{\left(k\right)}\mathcal{L}^{\left(k\right)} - 4\pi F_{ab}F^{ab}\Bigr)\,,
\label{eq:Loveaction}
\end{equation}
where $K=\left[\frac{d-1}{2}\right]$, the  $\hat{\alpha}_{\left(k\right)}$ are the  Lovelock coupling constants, and $\mathcal{L}^{\text{\ensuremath{\left(k\right)}}}$ are the $2k$-dimensional Euler densities, given by 
$\mathcal{L}^{\left(k\right)}=\frac{1}{2^{k}}\,\delta_{c_{1}d_{1}\ldots c_{k}d_{k}}^{a_{1}b_{1}\ldots a_{k}b_{k}}R_{a_{1}b_{1}}^{\quad\;\, c_{1}d_{1}}\ldots R_{a_{k}b_{k}}^{\quad\;\, c_{k}d_{k}}\,,
$
with the  `generalized Kronecker delta function', \mbox{$\delta_{c_{1}d_{1}\ldots c_{k}d_{k}}^{a_{1}b_{1}\ldots a_{k}b_{k}}=\left(2k\right)!\delta^{c_{1}}_{[a_{1}}\delta^{d_{1}}_{b_{1}}\ldots\delta^{c_{k}}_{a_{k}}\delta^{d_{k}}_{b_{k}]}$},  totally antisymmetric in both sets of indices, and $R_{a_{k}b_{k}}^{\quad\;\: c_{k}d_{k}}$ the Riemann tensor. 
In what follows we always take all the Lovelock couplings to be positive,  
identify the bare (negative) cosmological {constant $\Lambda=-\hat{\alpha}_{0}/2=-8\pi P$, and set $\hat \alpha_1=1$.} 

It is believed that in Lovelock gravity, the entropy is no longer given by  one quarter of the horizon area, but rather reads~\cite{Jacobson:1993xs,Wald:1994}
\begin{equation}\label{SJacobson}
S_0= \frac{1}{4} \sum_k \hat{\alpha}_{k} {\cal A}^{(k)}\,,\quad   {\cal A}^{(k)}  = k\int_{\mathcal{H}} \sqrt{\sigma}{\mathcal{L}}^{(k-1)}\,.
\end{equation}
Here, $\sigma$ denotes the determinant of $\sigma_{ab}$,  the induced metric on the black hole horizon ${\mathcal{H}}$, and the Lovelock terms ${\mathcal{L}}^{(k-1)}$ are evaluated on that surface.

The charged spherically symmetric AdS Lovelock black holes take the following form \cite{Boulware:1985wk, Wheeler:1985nh} (see also \cite{Hull:2021bry} for a more general case):
\be\label{solution}
ds^{2} = -f dt^{2}+\frac{dr^2}{f}+r^{2}d\Omega_{d-2}^{2}\,,\quad 
F= \frac{Q}{r^{d-2}} dt\wedge dr\,,
\ee
where $d\Omega_{d-2}^{2}$ denotes the line element of a $\left( d-2 \right)$-dimensional  space of constant curvature $\kappa(d-2)(d-3)$, with  $\kappa=+1,0,-1$ for spherical, flat, and hyperbolic  geometries respectively of finite   volume   $\Sigma_{d-2}$, the latter two cases being compact via identification, e.g. 
\cite{Aminneborg:1996iz,Mann:1997iz}. After integration, the Lovelock equations reduce to the following polynomial equation for $f$ 
\ba \label{eq:poly}
{\cal P}\left(f\right)&=&\sum_{k=0}^{K}\alpha_{k} \left(\frac{\kappa-f}{r^2}\right)^{k}=\frac{16\pi M}{(d-2)\Sigma_{d-2}^{(\kappa)}r^{d-1}}\nonumber\\
&&-\frac{8\pi Q^2}{ (d-2)(d-3)}\frac{1}{r^{2d-4}}\,,
\ea
where  $M$ stands for the ADM mass of the black hole, $Q$ is the electric charge, and
\ba
\alpha_{0}&=&\frac{\hat{\alpha}_{(0)}}{\left(d-1\right)\left(d-2\right)}=\frac{16\pi P}{\left(d-1\right)\left(d-2\right)}\,,
\quad {{\alpha}_{1}=1}\,,\nonumber\\
\alpha_{k}&=&\hat \alpha_{(k)}\prod_{n=3}^{2k}\left(d-n\right)  {\quad\mbox{for}\quad  k\geq2}
\ea
are the rescaled Lovelock couplings.

The `standard' thermodynamic quantities are then given by \cite{Cai:2003kt}:
\ba
M&=&\frac{\Sigma_{d-2}^{(\kappa)}\left(d-2\right)}{16\pi}\sum_{k=0}^{K}\alpha_{k}\kappa^kr_+^{d-1-2k}+\frac{\Sigma_{d-2}^{(\kappa)}}{2(d-3)}\frac{Q^2}{r_+^{d-3}}\,,\nonumber\\
T_0&=&  \frac{\vert f^\prime(r_+)\vert}{4\pi} =
\frac{T_{\text{mod}}}{\Delta}\,,\quad \Phi=\frac{\Sigma_{d-2}^{(\kappa)}Q}{(d-3)r_+^{d-3}}\,,\nonumber
\label{T}\\
S_0&=&S+\frac{\Sigma_{d-2}^{(\kappa)}\left(d-2\right)}{4}\sum_{k=2}^{K}\frac{k\kappa^{k-1}\alpha_{k}r_+^{d-2k}}{d-2k}\,, \quad 
 \label{eq:Entropy}
\ea
where
\ba
T_{\text{mod}}&=&\frac{1}{4\pi r_+}\Biggl[\sum_{k=0}^{K}\kappa\alpha_k(d\!-\!2k\!-\!1)\Bigl(\frac{\kappa}{r_+^2}\Bigr)^{k-1}\!\!\! \nonumber\\
&&-\frac{8\pi Q^2}{(d-2)r_+^{2(d-3)}}\Biggr]\,,\nonumber\\ \label{T0}
S&=& \frac{\Sigma_{d-2}^{(\kappa)}\alpha_1 r_+^{d-2}}{4}\,.
\ea
That is, the leading term in the expression for $S_0$ is one-quarter the horizon area, $S$, while the other terms come from higher-curvature contributions. We also have  
\be
\Delta=\sum_{k=1}^{K}k\alpha_{k}\left(\kappa/ r_{+}^{2}\right)^{k-1}\,. 
\ee
These quantities obey the following first law and Smarr relations
\cite{Jacobson:1993xs, Kastor:2010gq}:
\ba
\delta M&=&T_0\delta S_0{+}\frac{1}{16\pi}\sum_{k}\Psi^{\left(k\right)}\delta{\alpha}_{\left(k\right)}
+\Phi\delta Q\,,\label{first}\\
M&=&\frac{d-2}{d-3}T_0S_0+\sum_{k}\frac{2\left(k-1\right)}{d-3}\frac{ \Psi^{\left(k\right)}{\alpha}_{\left(k\right)}}{16\pi}+\Phi Q\,,\qquad
\label{Smarr}
\ea
where 
\be\label{V}
V=\frac{16\pi \Psi^{(0)}}{(d-1)(d-2)}=\frac{\Sigma_{d-2}^{(\kappa)}r_+^{d-1}}{d-1}\,,
\ee
is the thermodynamic volume and expressions for the remaining potentials $\Psi^{\left(k\right)}$ can be found in \cite{Frassino:2014pha}. 

As an alternative to the standard thermodynamics above, one may easily check that taking $T_{\text{mod}}$ as the black hole temperature, and identifying the entropy with the horizon area, yields also the consistent thermodynamic laws \eqref{first}--\eqref{Smarr}, where 
\be
T_0\to T_{\text{mod}}=T_0\Delta\,,\quad S_0\to S\,, 
\ee
with the same thermodynamic volume $V$, and modified $\Psi^{(k)}_{\text{mod}}$ ($k\geq 2)$, which now take a very simple form:
\be
\Psi^{(k)}_{\text{mod}}={\left(d-2\right)\Sigma_{d-2}^{(\kappa)}}
\kappa^{k}r_+^{d-2k-1}.
\ee
In other words, upon modifying the black hole temperature, by $T_0\to T_0\Delta$, we can preserve the area law. Note that the factor $\Delta$ precisely reduces to that found for the 4D Gauss-Bonnet gravity upon taking the limit to $d=4$ dimensions.  
Note also that upon identifying the entropy according to the area law, for $\kappa=-1$, $S$ may no longer become negative, as is the case with the entropy \eqref{SJacobson}.

Notably, the modified temperature $T_{\text{mod}}$ is again constant on the horizon and the zeroth law of black hole mechanics continues to hold. For the solutions of Lovelock gravity  possessing a Killing vector $\xi^{a}$, we even found a way to modify the action by a boundary term to recover the area law entropy and modified temperature. In particular, for the $(k=2)$ case of the Gauss-Bonnet gravity such a term reads 
\be
\frac{\alpha_2}{8\pi}\int\nabla_{a}\left(\frac{\partial\mathcal{G}}{\partial R_{abcd}}\xi_{b}\nabla_{c}\xi_{d}\right)\sqrt{-\mathfrak{g}}\text{d}^d\!x\,, 
\ee
see appendix~\ref{AppA} for more details.

On the other hand, it is not clear at the moment whether the second law remains valid for this choice of entropy. Nevertheless, the Hawking area increase theorem~\cite{Hawking:1971}, one of the classical underpinnings of the identification of entropy with area, continues to hold in Lovelock gravity assuming the null curvature condition, i.e., $R_{ab}k^{a}k^{b}\ge 0$, for any null, future-pointing vector $k^{\mu}$. This condition is also required to prevent the presence of certain pathological behavior such as existence of closed timelike curves in modified theories of gravity.

\section{Discussion}

In this work, we have shown that, contrary to the claims in the literature, the area law can prevail for black holes in the recently formulated scalar-tensor 4D Gauss--Bonnet gravity. The validity of the first law then implies that the black hole temperature is accordingly modified. We have also noted that similar modification preserves the area law for black holes in higher-dimensional Lovelock theories, giving rise to `significantly simplified' thermodynamic quantities.

The price paid for preserving the Bekenstein's law is the modified black hole temperature. This is no longer given by the surface gravity but picks up an additional (constant on the horizon) factor. For 4D Einstein--Gauss--Bonnet gravity such a modification simply originates from adding a covariant boundary term ensuring the shift symmetry of the action. 
The temperature can then be calculated by the Brown--York procedure. Viewing this 4D theory as a four-dimensional limit of  a (purely metric) higher-dimensional Einstein--Gauss--Bonnet gravity provides a hint for modifying the thermodynamics of the latter. In fact, action of any Lovelock theory can be modified by a boundary term to recover the area law for entropy, although the procedure (so far) works only for spacetimes with a Killing symmetry.

However, what would be the physical origin of such modified temperature? In the Horndeski gravity it was argued \cite{Hajian:2020dcq} that the modification of the temperature arises from the  fact that, in the presence of the non-minimally coupled scalar field, the gravitons (that are supposed to dominate Hawking radiation) no longer propagate along the background geometry but rather follow the effective metric. Natural question is whether a similar explanation can also be given in the case of the 4D Gauss Bonnet gravity or even in the (purely metric) Lovelock theory. It is established that the speed of propagation of gravitons is modified, due to the higher-curvature corrections, in these theories \cite{Izumi:2014,Papallo:2015,Benakli:2016}. Notably, while the Hawking original calculation does not refer to gravitational dynamics in any way, it is dependent on the speed of propagation of the emitted particles. A striking example of this dependence can be found in the acoustic analogues of the Hawking effect, in which the relatively low speed of the emitted acoustic modes increases the Hawking temperature to measurable values \cite{Barcelo:2011}.

A similar effect has also been observed in theories of non-linear electrodynamics, where the two degrees of freedom suffer from birefringence and lead to distinct propagation speeds of non-linear photons, e.g. \cite{plebanski1970lectures}. Could this have an impact on black hole thermodynamics in such theories?

A potential problem with computing the Hawking temperature from graviton propagation exists. While the gravitons dominate the Hawking radiation for rapidly rotating black holes, for spherical black holes the dominant contribution comes from photons \cite{Page:1976}. Then, it remains unclear why one should single out the graviton propagation speed as the decisive one for the Hawking evaporation. Nevertheless, for a vacuum classical black hole spacetime, gravity is the only field present. Perhaps this somehow gives gravitons a privileged position. In any case, the issue requires further attention. 

Interestingly, neither the Wald procedure, nor the `standard' Euclidean action calculation are in a good position to determine the black hole temperature. Instead, they only fix an expression for the product of temperature and entropy. The black hole temperature is often derived by the Euclidean trick, demanding regularity of the Wick-rotated geometry. This may cause an impression that black hole temperature is a local quantity that can be derived from the properties of the horizon. However, this is not entirely true, as it is also affected by the redshift that the Hawking radiation experiences on its way to the asymptotic region (encoded in the normalization of the Killing horizon generator at infinity), and thence in principle depends on the matter content between the horizon and infinity. These nonlocal effects provide another possible explanation for the temperature modifications besides the modified speed of gravitons.

The Brown--York grand canonical ensemble procedure~\cite{Braden:1990hw, Brown:1992bq}, which takes into account this nonlocality, can recover modified temperature (depending on the boundary terms one adds to the action).  Although such modifications do not seem to occur in the presence of minimally coupled matter fields \cite{Bardeen:1973}, the example of Horndeski theories suggests that they can occur in the presence of non-minimal coupling (i.e., when the matter fields modify the graviton propagation, as is the case for the effective energy-momentum tensor of Lovelock gravity). Could the temperature also be affected by additional (higher-curvature) non-linearities in the gravitational field itself?

Let us also stress that although we have presented a  rather convincing argument for the preservation of the area law in the case of the 4D Gauss--Bonnet gravity, the alternative formulation of the thermodynamics of higher-dimensional Lovelock black holes is at this point just an observation. Moreover, this observation has only been tested for `spherical black hole spacetimes' (with trivial extension to a charged case) whose thermodynamic first law is of `cohomogeneity one' (depends only on one non-trivial thermodynamic parameter). In such a case, it is always possible to fix one of the conjugate entropy-temperature pair and `calculate' the second one by demanding the first law. This is however, no longer true in the presence of rotation. It will thus be a non-trivial test whether the area law may also prevail for (spherical) rotating black holes in Lovelock gravity, once these are found. This, together with a check of the second law, will ultimately allow us to decide whether the area law may remain  valid in any Lovelock gravity.

Finally, our findings uncover a crucial dependence of derived thermodynamic quantities on the presence of certain boundary terms in the action. While such terms cannot modify the dynamics of the theory, the physical implications on the interpretation of the corresponding thermodynamic ensembles remain to be clarified.

\subsection*{Acknowledgements}
We would like to thank Roberto Emparan, Robie Hennigar, Shahin Sheikh Jabbari, Robert Mann, and Mohammad Hassan Vahidinia for very useful comments on this work.  
D.K. is grateful for support from GA{\v C}R 23-07457S grant of the Czech Science Foundation. M.L. is supported by the Charles University Grant Agency project No. GAUK 90123.

\appendix

\section{Area law for 5D Einstein-Gauss-Bonnet gravity}\label{AppA}

In this appendix we show that both the modified temperature and the Bekenstein's area law for 5D Einstein-Gauss-Bonnet gravity can be derived from the standard Euclidean action calculation à la Brown-York~\cite{BrownYork}. It provides a very physical method standardly used in black hole thermodynamics that allows for explicit derivation of the (modified) black hole temperature, and is completely independent from the covariant phase space approach and the Wald entropy formula discussed in the main text.

The action of 5D Einstein--Gauss--Bonnet gravity reads (setting the cosmological constant $\alpha_0=0$ for simplicity)
\begin{equation}
	I_{1}=\frac{1}{16\pi}\int\left(R+\alpha\mathcal{G}\right)\sqrt{-\mathfrak{g}}\text{d}^5\!x,
\end{equation}
where $\mathcal{G}=R^2-4R_{ab}R^{ab}+R_{abcd}R^{abcd}$. Both Wald formula and Euclidean canonical ensemble calculations lead to standard Hawking temperature and modified entropy for this action.

In 4D scalar-tensor Einstein--Gauss--Bonnet gravity, we recovered the area law entropy using that $\mathcal{G}=\nabla_{a}\mathcal{G}^{a}$ in 4D and adding a covariant boundary term $\alpha\nabla_{a}\left(\phi\mathcal{G}^{a}\right)/16\pi$ to the action. Since in 5D $\mathcal{G}$  is no longer a total divergence, we cannot use a similar trick here. Nevertheless, suppose we limit ourselves to spacetimes possessing some Killing vector $\xi^{a}$ (among other cases, this covers all stationary black hole solutions). Then, in 4D we have a simple expression for $\mathcal{G}^{a}$\cite{Padmanabhan:2011},\footnote{{An analogous expression holds for any Lovelock density in its critical dimension, just with $\mathcal{G}$ replaced by the density in question. Hence, the procedure to recover the area law outlined here for 5D Einstein--Gauss--Bonnet gravity in principle works for any Lovelock theory.}} 
\begin{equation}
	\mathcal{G}^{a}=-2\frac{\partial\mathcal{G}}{\partial R_{abcd}}\xi_{b}\nabla_{c}\xi_{d}\,.
\end{equation}
Remarkably, this expression is perfectly well defined even in 5D spacetimes with a Killing vector, although its divergence is no longer equal to $\mathcal{G}$. This allows us to consider the following action for 5D Einstein--Gauss--Bonnet gravity
\begin{equation}
	I_{2}=I_1+\frac{\alpha}{8\pi}\int\nabla_{a}\left(\frac{\partial\mathcal{G}}{\partial R_{abcd}}\xi_{b}\nabla_{c}\xi_{d}\right)\sqrt{-\mathfrak{g}}\text{d}^5\!x\,,
\end{equation}
which leads to the same equations of motion as $I_1$. However, it is easy to check that Wald entropy derived from this action reads
\begin{equation}
	S=\frac{\text{Area}}{4}\,,
\end{equation}
and we thus recover the area law. Then, to satisfy the first law of black hole mechanics, the temperature must be modified in the way we propose in the main text.

We can even define a Euclidean canonical ensemble for this metric and use it to compute the temperature directly. Consider a static, spherically symmetric Euclidean metric of the form
\begin{equation}
	\label{euc metric}
	\text{d}s^2=\left(b\left(y\right)\right)^2\text{d}\tau^2+\left(a\left(y\right)\right)^2\text{d}y^2+\left(r\left(y\right)\right)^2\text{d}\Omega_{3}^2
\end{equation}
where $\tau$ is the Euclidean time coordinate $\tau\in\left[0,2\pi\right)$, and coordinate $y\in\left[0,1\right]$ is chosen so that $r\left(0\right)=r_+$ corresponds to the black hole event horizon and $r\left(1\right)=r_{\text{b}}>r_+$ to the artificial boundary of the spacetime with topology $S^{1}\times S^{3}$. For the geometry to be regular at the horizon, we must have $b\left(0\right)=0$ and $\left(b'/a\right)_{y=0}=1$~\cite{BrownYork}. The inverse temperature $\beta$ measured by a static observer on the boundary is given by the proper length of its $S^{1}$ component, i.e., $\beta=2\pi b\left(1\right)$~\cite{BrownYork}. The action $I_2$ evaluated for this metric reads
\begin{eqnarray}
	I_{2}&=&-\frac{\pi^2}{2}\int_{0}^{1}\text{d}y\bigg[3rab+\frac{3r^2r'b'}{a}+\frac{3rr'^2b}{a} \\
	&&-\bigg(\frac{r^3b'}{a}+\frac{3r^2r'b}{a}\bigg)'+4\alpha\left(\frac{br'^3}{a^3}-3\frac{br'}{a}\right)' \nonumber\\
	&&+4\alpha\bigg(\frac{r'^3b'}{a^3}-3\frac{r'b'}{a}\bigg)\bigg]\,,
\end{eqnarray}
where $'$ denotes a derivative with respect to $y$ and we already carried out integration over $\text{d}\tau\text{d}\Omega_3$. To set Dirichlet boundary conditions on the boundary at $y=1$, we need to add the following boundary term to the action
\begin{align}
	\nonumber I_{\text{b}}=&\frac{\pi^2}{2}\bigg[-\left(\frac{r^3b'}{a}+\frac{3r^2r'b}{a}\right)+3r^2b \\
	&+4\alpha\left(\frac{br'^3}{a^3}-3\frac{br'}{a}\right)+8\alpha b\bigg]_{y=1}.
\end{align}
After some straightforward manipulations, the total action, $I=I_2+I_{\text{b}}$, reads
\begin{eqnarray}
	I&=&-\frac{\pi^2}{2}\int_{0}^{1}\text{d}y\bigg[3rab+\frac{3r^2r'b'}{a}+\frac{3rr'^2b}{a}+\frac{4\alpha r'^3b'}{a^3} \nonumber\\
	&&-\frac{12\alpha r'b'}{a}\bigg]+\frac{\pi^2}{2}\left[3r^2 b+8\alpha b\right]_{y=1} \\
	&&-\frac{\pi^2}{2}\bigg[\bigg(\frac{r^3b'}{a}+\frac{3r^2r'b}{a}\bigg)-4\alpha\left(\frac{br'^3}{a^3}-3\frac{br'}{a}\right)\bigg]_{y=0}\,.\nonumber
\end{eqnarray}
The constrain equation obtained by varying the action with respect to $b$ fixes
\begin{equation}
	\zeta\left(r\right)\equiv	\frac{r'\left(y\right)}{a\left(y\right)}=\sqrt{1+\frac{r^2}{2\alpha}\left(1-\sqrt{1+\frac{4\alpha r_+^2}{r^4}\bigl(1+\frac{\alpha}{r_+^2}\bigr)}\right)}\,.
\end{equation}
Plugging this back into the action and integrating finally yields
\begin{equation}
	I=\frac{3\pi}{4}r_{\text{b}}^2\beta\bigl(1-\zeta\left(r_{\text{b}}\right)\bigr)+\pi\alpha\beta\bigl(2+3\zeta\left(r_{\text{b}}\right)-\zeta^3(r_{\text{b}})\bigr)\,,
\end{equation}
where we recall $\beta=2\pi b\left(1\right)$. Stationary points of this action are given by condition $\partial I/\partial r_+=0$, which allows us to compute the equilibrium value of inverse temperature $\beta$. In the limit of $r_{\text{b}}\to\infty$ (i.e., for the stationary observers at asymptotic infinity), it reads
\begin{equation}
	\beta=2\pi r_{+},
\end{equation}
which is precisely the inverse of the modified temperature we proposed in~\eqref{T0}. Finally, entropy can be obtained by the standard formula
\begin{equation}
	S=\beta\partial_{\beta}I-I=\frac{\pi^2r_+^3}{2}=\frac{\text{Area}}{4}\,,
\end{equation}
in agreement with the area law.

The main drawback of the described procedure is that it works only in spacetimes with a Killing vector. Nevertheless, this covers the case of all stationary black hole solutions. Moreover, it might be possible to exploit the link between 4D scalar-tensor and 5D Einstein-Gauss-Bonnet gravities further, and obtain a fully general version of the boundary term we consider here.

\bibliography{Databaze2}

\end{document}